\documentclass[10pt,conference]{IEEEtran}

\IEEEoverridecommandlockouts
\usepackage{amsmath,amssymb,amsfonts}
\usepackage{mathptmx} % This is Times font
\usepackage{fancyhdr}
\usepackage[normalem]{ulem}
\usepackage{graphicx}
\usepackage{textcomp}
\usepackage{xcolor}
\usepackage[hyphens]{url}
\usepackage{comment}
\usepackage[sort,nocompress,noadjust]{cite}

\usepackage{algorithm}% http://ctan.org/pkg/algorithm
\usepackage{algpseudocode}% http://ctan.org/pkg/algorithmicx
\usepackage[skip=0pt,font=smaller]{subcaption}% http://ctan.org/pkg/caption
\usepackage{array}
\usepackage{float}
\usepackage[bookmarks=true,breaklinks=true,letterpaper=true,colorlinks,citecolor=blue,linkcolor=blue,urlcolor=blue]{hyperref}
\usepackage{paralist}
\usepackage{tikz}
\newcommand{\ballnumber}[1]{\tikz[baseline=(myanchor.base)] \node[circle,fill=.,inner sep=0.2pt] (myanchor) {\color{-.}\footnotesize #1};}
\newcommand{\ballblue}[1]{\tikz[baseline=(myanchor.base)] \node[inner color=white,shape=circle,draw,fill=.,color=blue,inner sep=0.1pt] (myanchor) {\color{black}\footnotesize #1};}

\newfloat{Algorithm}{htbp}{loa}

\def\BibTeX{{\rm B\kern-.05em{\sc i\kern-.025em b}\kern-.08em
    T\kern-.1667em\lower.7ex\hbox{E}\kern-.125emX}}

\newcolumntype{A}{>{\centering\arraybackslash}m{2.2cm}}
\newcolumntype{B}{>{\centering\arraybackslash}m{1.85cm}}
\newcolumntype{E}{>{\centering\arraybackslash}m{1.9cm}}
\newcolumntype{F}{>{\centering\arraybackslash}m{2.1cm}}

\newcolumntype{G}{>{\centering\arraybackslash}m{2.2cm}}
\newcolumntype{H}{>{\centering\arraybackslash}m{2.2cm}}
\newcolumntype{J}{>{\centering\arraybackslash}m{1.6cm}}

\title{On Consistency for Bulk-Bitwise Processing-in-Memory}

\author{\IEEEauthorblockN{Ben Perach \qquad Ronny Ronen \qquad Shahar Kvatinsky}
\IEEEauthorblockA{\textit{The Andrew and Erna Viterbi Faculty of Electrical \& Computer Engineering} \\
\textit{Technion -- Israel Institute of Technology}\\
Haifa, Israel \\
benperach@campus.technion.ac.il \quad ronny.ronen@ef.technion.ac.il \quad shahar@ee.technion.ac.il}
}
 
\newcommand\blfootnote[1]{%
  \begingroup
  \renewcommand\thefootnote{}\footnote{#1}%
  \addtocounter{footnote}{-1}%
  \endgroup
}

\begin{document}
\maketitle
\thispagestyle{plain}
\pagestyle{plain}

%%%%%% -- PAPER CONTENT STARTS-- %%%%%%%%

\begin{abstract}

Processing-in-memory (PIM) architectures allow software to explicitly initiate computation in the memory. This effectively makes PIM operations a new class of memory operations, alongside standard memory operations (\textit{e.g.}, load, store). For software correctness, it is crucial to have ordering rules for a PIM operation with other PIM operations and other memory operations, \textit{i.e.}, a consistency model that takes into account PIM operations is vital. To the best of our knowledge, little attention to PIM operation consistency has been given in existing works.
In this paper, we focus on a specific PIM approach, named bulk-bitwise PIM. In bulk-bitwise PIM, large bitwise operations are performed directly and stored in the memory array. We show that previous solutions for the related topic of maintaining coherency of bulk-bitwise PIM have broken the host native consistency model and prevent any guaranteed correctness.
As a solution, we propose and evaluate four consistency models for bulk-bitwise PIM, from strict to relaxed. Our designs also preserve coherency between PIM and the host processor. Evaluating the proposed designs' performance with a gem5 simulation, using the YCSB short-range scan benchmark and TPC-H queries, shows that the run time overhead of guaranteeing correctness is at most $6\%$, and in many cases the run time is even improved. The hardware overhead of our design is less than $0.22\%$.

\end{abstract}

\section{Introduction}
\label{sec:Intro}
\blfootnote{This work was supported by the European Research Council through the European Union's Horizon 2020 Research and Innovation Programme under Grant 757259 and through the European Union's Horizon Europe Research and Innovation Programme under Grant 101069336.}
In recent processing-in-memory (PIM) architectures, where the memory module does not only hold data but also processes it, PIM operations are explicitly initiated by software (\emph{e.g.}, by a dedicated instruction in the host processor~\cite{PIMDB,AMBIT,SIMDRAM,OrderLight,Ahn2015,LazyPIM}). When initiated, these PIM operations are sent from the host core to the memory, thereby becoming a new class of memory operations alongside standard memory operations such as load and store. These new operations require re-investigation of the consistency model for hosts using PIM operations. Questions about reordering of PIM operations between PIM and other memory operations must be raised. These issues are important, as without clear ordering rules, it is hard to reason about program correctness~\cite{PrimerBook}. Despite its importance, the issue of a consistency model for PIM has been completely ignored in PIM architecture works~\cite{PIMDB,AMBIT,SIMDRAM,RACER,MOUSE,Ahn2015,Ghose2018,FloatPIM,LazyPIM,CONCEPT}.

PIM techniques can be categorized according to the location of the processing units, the granularity of the PIM operations, and the arbitration of the PIM computations and memory accesses~\cite{OrderLight,Reuben2017}. Each PIM category exhibits different characteristics, suits different application classes, and requires different system-level support. In this paper, we focus specifically on \textit{bulk-bitwise PIM}~\cite{PIMDB,AMBIT,SIMDRAM,CONCEPT,RACER,Pinatubo,MOUSE} and discuss the consistency of such PIM techniques. 

In bulk-bitwise PIM, the memory array serves both as the storage medium and as the processing unit, performing the computation on its stored data and writing the result directly to the memory cells within the array. The bulk-bitwise operations performed within the array are bitwise row-to-row or column-to-column logic operations, potentially performed on numerous memory arrays simultaneously. These characteristics have two important implications on the computing system: a) bulk-bitwise PIM operations are immediately committed to the system state during PIM execution, and b) a single bulk-bitwise PIM operation may change a large memory section. Regarding the consistency model and its implementation, these implications suggest that bulk-bitwise PIM operations cannot be speculatively executed and reverted on order violation and that the ordering of PIM operations needs to be enforced on a large address range. These implications do not necessarily apply to other PIM techniques~\cite{Ahn2015,LazyPIM,Ghose2018,OrderLight,Reuben2017}, which might require different consistency models and different solutions.

\begin{figure}[!t]
\noindent\begin{minipage}{.2\textwidth}
\centering
\begin{algorithmic}
    \State \textbf{\underline{Example Code:}}
    \State Write(A)   \textit{// step }\ballnumber{1}
    \State MemFence
    \State Write(B)   \textit{// step }\ballnumber{2}
    \State MemFence
    \State Flush(A)   \textit{// step }\ballnumber{3}
    \State Flush(B)   \textit{// step }\ballnumber{4}
    \State MemFence
    \State PIM op     \textit{// step }\ballnumber{6}
\end{algorithmic}
\end{minipage}%
\hspace{10pt}%
\begin{minipage}{.24\textwidth}
  \centering
  \includegraphics[width=\textwidth]{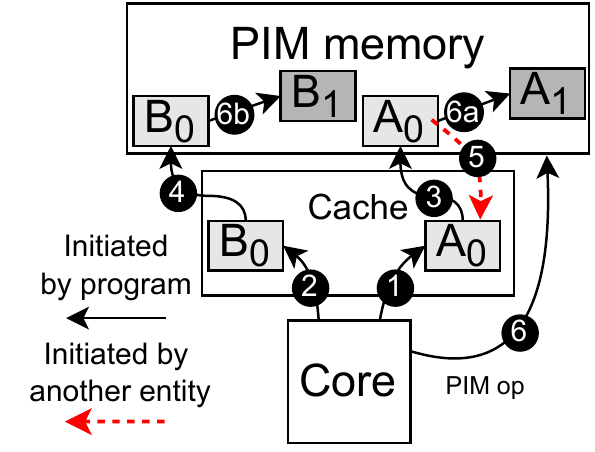}
\end{minipage}
\captionof{figure}{Example of a code using a PIM operation with a possible reordering of memory operations, resulting in a cyclic order; see the example described in Section~\ref{sec:Intro}. Execution steps are marked as comments in the code and dark circles. \textit{MemFence} is a standard memory fence instruction, indicating strict order between operations before and after the fence~\cite{OrderLight,PrimerBook}. The cyclic order in the example breaks the \textit{MemFence} order guarantees.}
\label{fig:example}
\end{figure}

Some prior works refer to the related topic of coherency between the PIM module and the host caches~\cite{SIMDRAM,PIMDB,AMBIT}. Consistency and coherency are tightly coupled when PIM is considered, as loads and stores are handled in the host processor caches while PIM operations go to the memory, potentially reordering PIM operations with loads and stores. Designing coherency solutions for PIM should be done with care, as flawed ones can result in problematic behavior. For instance, in~\cite{SIMDRAM,PIMDB}, PIM operations do not interact with the host processor caches and the coherency between the PIM module and caches is the responsibility of the software (by using cache flushes). For such a PIM operation mechanism, consider the scenario depicted in Fig.~\ref{fig:example}. A thread, running the example code in Fig.~\ref{fig:example}, writes value $A_0$ to address $A$ (Fig.~\ref{fig:example}\ballnumber{1}) and then writes value $B_0$ to address $B$ (Fig.~\ref{fig:example}\ballnumber{2}). Later, both addresses are flushed to memory (Fig.~\ref{fig:example}\ballnumber{3}\ballnumber{4}) before issuing a PIM operation (Fig.~\ref{fig:example}\ballnumber{6}) that will change their values to $A_1$ and $B_1$ (Fig.~\ref{fig:example}\ballnumber{6a}\ballnumber{6b}). Between the flush and the PIM operation, however, $A$ is read from the memory to a cache (Fig.~\ref{fig:example}\ballnumber{5}) with the value $A_0$ (\emph{e.g.}, by another thread or by a prefetcher). As a result, read operations to $A$ after the PIM operation will result in a cache hit and the old value of $A_0$.
Although the software is written properly (\textit{i.e.}, all related addresses are flushed from the cache by the software), the result is still incorrect without a proper consistency model.

The above scenario is problematic for an additional reason.
A \textbf{cyclic ordering} without a well-defined happen-before relation exists in this scenario, allowing the same thread to see different orders of events regardless of the host processor's native consistency and software enforcement.
A cyclic ordering can be formed by the following sequence of operations: (1) reading addresses $A$ and $B$ to validate that $Write(A)$ happened before $Write(B)$, (2) reading $B$ twice and getting $B_0$ and then $B_1$ to validate that $Write(B)$ is before $PIM op$, and (3) reading $B$, getting $B_1$, and then reading $A$, getting $A_0$ (from the cache) to validate that $PIM op$ is before $Write(A)$. Hence, $Write(A)$ is before $Write(B)$, $Write(B)$ is before $PIM op$, and $PIM op$ is before $Write(A)$, \textit{i.e.}, $Write(A)$ is before itself, forming a cycle.
Furthermore, the ordering of $Write(A)$ and $Write(B)$ is not well-defined, \textbf{breaking the ordering rules of the host} specified explicitly by the \textit{MemFence} instructions~\cite{PrimerBook}. Changing the host ordering rules by the additional PIM operation breaks the software correctness relying on these rules. 

In the above example, the problem comes from the non-atomicity of the PIM operation and cache flushes. This non-atomicity enables relevant addresses to be brought back to the host caches, making the PIM memory non-coherent with the host caches, and violating the host processor ordering rules for loads and stores. Any ordering guarantees that include PIM operations require the atomicity of PIM operations and cache flushes on the relevant addresses.

In this paper, we propose and evaluate different options for incorporating PIM operations into a consistency model. We offer hardware solutions to enforce the proposed consistency models, including the atomicity of the PIM operations and cache flushes. Our solutions are evaluated using the gem5 simulator environment~\cite{gem5} and workloads for the YCSB~\cite{YCSB} and TPC-H~\cite{TPCH} benchmarks.

In summary, this paper makes the following contributions:
\begin{compactitem}
    \item We propose four consistency models for bulk-bitwise PIM operations and suggest how to implement them.
    \item We design a hardware solution to enforce the atomicity of PIM operations and their required cache flushes.
    \item We evaluate our solutions using a gem5 simulation and database workloads (YCSB and TPC-H), showing that the run time overhead of guaranteeing correctness is at most $6\%$, and in many cases the run time is even improved.
    \item We show that in the context of bulk-bitwise PIM, relaxed consistency models do not necessarily execute faster than strict consistency models.
\end{compactitem}

\section{Background}
\subsection{Bulk-Bitwise PIM}
\label{subsec:background_PIM}
\begin{figure}[!t]
\centering
\includegraphics[width=0.5\columnwidth]{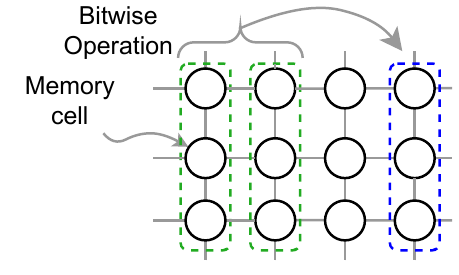}%
\caption{Illustration of a $3\times4$ memory array performing a simple bitwise column logic operation (\textit{e.g.}, NOR) per row. In each row, the operation inputs and output are the two left column cell values and the right column cell, marked by green and blue frames, respectively. Bitwise row operations are performed similarly.}
\label{fig:bitwise_logic}
\end{figure}

Bulk-bitwise PIM is a PIM technique characterized by the location of the PIM processing elements and their basic operation capabilities. The processing elements in bulk-bitwise PIM are the memory cells themselves and their periphery circuits (\textit{e.g.}, voltage drivers, sense amplifiers, decoders). Several technologies have been suggested for implementing such memory arrays, including DRAM~\cite{AMBIT,SIMDRAM} and emerging resistive technologies~\cite{MOUSE,Borghetti2010,Hoffer2020,MAGIC,Lyle2010} (often referred to as memristive stateful logic). All such technologies execute simple logic operations (\textit{e.g.}, AND, NOR, NOT) between one or more cells in the memory array and write the result into a memory cell, as shown in Fig.~\ref{fig:bitwise_logic}. These logic operations have the restriction that the input and output cells have to be on the same row or column. The same logic operation, however, can be performed in parallel on numerous cells that are aligned on the same columns or rows. Additionally, as a memory chip comprises many memory arrays, and a memory card comprises several memory chips, a memory card with such PIM capabilities can concurrently operate on many arrays, achieving substantial computational throughput, \textit{i.e.}, bulk-bitwise operations.

To perform more complex operations (\textit{e.g.}, addition, multiplication), the basic array logic operations are performed multiple times, during which the array cannot be accessed, and requires additional memory cells to hold intermediate values~\cite{SIMPLER,SIMDRAM}. Hence, performing a complex operation may implicitly change more cells than just the designated output cells. To support these complex operations at the memory array level, an additional control logic is added to generate the sequence of basic operations~\cite{PIMDB,SIMDRAM,RACER,SIMPLER} and manage the additional cells. Due to area and control constraints, the control logic is often shared between several arrays, and performs the same operation in parallel on all shared arrays.

Since bulk-bitwise PIM operations change memory cell values, using bulk-bitwise PIM modules as the main memory inherently changes the visible state of the system. Therefore, when a command to perform a bulk-bitwise PIM operation is sent from a host processor, the host must issue the command only when it reaches the pipeline commit stage. In that respect, bulk-bitwise PIM operations are similar to store instructions. Nevertheless, unlike a store instruction, a PIM operation is sent directly to the memory and not to the cache hierarchy. Also unlike a store command, since intermediate results are stored within the memory cells, from the host point of view, a PIM operation may change unexpected memory regions. Furthermore, the fact that the main memory is a different module, separated from the host chip, allows different PIM modules with different technologies and instruction sets to connect to the same host. Different PIM modules might mean different encoding for a PIM operation, different complex operations, and different implicit changes of memory cells due to these complex operations. Hence, for a host supporting multiple different PIM modules, the host hardware cannot easily know what addresses are affected by a PIM operation.

Supporting virtual memory is possible with bulk-bitwise PIM. Recent work~\cite{PIMDB} has shown that bulk-bitwise PIM operations can be restricted to work within the boundaries of a single memory huge-page~\cite{HWSW_IF}. This allows user-level code to issue PIM operation commands over virtual memory using the standard memory translation procedure. PIM huge-pages can be allocated to the extent that the PIM memory capacity allows, where the computations within different huge-pages are independent. To perform the same computation on several pages, a PIM operation should be sent to each page separately.

\subsection{Consistency Models}
\label{subsec:background_consistency}
In a shared memory system, where multiple threads concurrently access the same memory region, the order of these accesses \textit{must} be well defined~\cite{PrimerBook}. Using the ordering of the memory accesses, a multi-threaded program can be proven to correctly or incorrectly execute the desired task; otherwise, the correct execution of a program cannot be guaranteed. For instance, if a read and a write to the same memory address can be performed in any order, we cannot know if the read result will be the value before or after the write operation. The possible ordering of memory accesses, or the memory ordering rules for a system, is referred to as the \textit{consistency model} of that system~\cite{PrimerBook}.

When reordering of memory operations is discussed, the reference is to the difference between two order relations, the \textit{program order} relation of the issuing thread and the \textit{memory order} relation of another thread, each defining a happen-before relation. A thread's program order is the order of memory operations appearing in the program code of the thread. A thread's memory order is the order of all memory operations in the system as seen by the thread.
Memory operations are reordered if (1) they appear in the program order of thread 1, and (2) they appear in a different order in the memory order of thread 2. 

Numerous consistency models exist in modern processors~\cite{PrimerBook}. These can be ranked by how strict or relaxed their ordering rules are. Whereas stricter models define more rules that must be obeyed, relaxed models allow more reordering possibilities. To enforce order where reorder is possible, consistency models supply additional operations, \textit{e.g.}, atomic operations (also referred to as read-modify-write operations~\cite{Bharghava2013}) and memory fences. Atomic operations read a value from memory, modify it, and write it back to the memory such that these operations are consecutive and cannot be reordered with any other memory operation. Memory fence operations ensure that memory operations before and after the fence (in program order) cannot be reordered. Due to the possible reordering of memory operations, some subtle and unexpected behaviors can occur when consistency models become more relaxed~\cite{Chakraborty2019}, making it hard to program with relaxed models. The motivation for having more relaxed consistency models is to make the hardware more performant by allowing concurrent operations of more memory requests. 

\section{Consistency Models For Bulk-Bitwise PIM}
\label{sec:pim_consistency}
This section presents four possible consistency models for PIM operations, from the strictest to the most relaxed. These models extend, without violating, the existing host processor consistency model by specifying the ordering rules involving PIM operations. Ordering rules not involving PIM operations of existing models are not modified. Implementations are discussed in Section~\ref{sec:consistency_support}. 

Before we present the proposed consistency models, we define our PIM operation model. A PIM operation, in the context of this paper, referred to as a \textit{PIM op}, is a memory operation issued by a thread to a specific memory address range.
We call this range of addresses the \textit{scope} of the PIM op. The PIM op does not necessarily use or modify all addresses in its scope; it is, however, limited to operating only within these addresses. Furthermore, the PIM memory is partitioned into a fixed set of scopes, each with a fixed address range with no addresses overlapping between scopes. The fixed set of scopes and scope sizes (which can range from a cache-line~\cite{Ahn2015} to the entire memory~\cite{AMBIT,CONCEPT}) are architectural values, defined by the system being used. PIM ops can only be issued to scopes from this fixed set. If data in multiple scopes require the same processing, the required PIM ops should be duplicated for each scope. When a PIM op is issued, its scope identity must be available to the host hardware (\textit{e.g.}, the address of the PIM op), providing the host information where to route the PIM op and what addresses might be affected by the PIM op. The scope identification is implementation depended, for example, in~\cite{PIMDB}, 1GB huge-pages are used as scopes and are identified by the PIM op address. Knowing the scope of a PIM op, not the exact addresses that are read and written by the operation, forms an abstraction for the host hardware. The scope abstraction enables the same host to connect with different PIM modules, each possibly utilizing different PIM instruction sets. 

We now describe four potential consistency models from the strictest to the most relaxed. 

\textbf{Atomic model.} For our first proposed consistency model, we observe that bulk-bitwise PIM ops, when reaching the memory, are performed atomically. By atomically, we mean in the sense that once the PIM op starts execution, the memory array is occupied until the operation is complete, not allowing other PIM ops, reads, or writes to be executed. As a PIM op can be modeled by many loads and stores, it seems intuitive to model the behavior of a PIM op as multiple loads and stores performed atomically, \textit{i.e.}, an atomic read-modify-write operation~\cite{PrimerBook} for a large address range. If we view a PIM op as an atomic operation for its scope, we get a consistency model for PIM ops where no memory operation from the same thread may be reordered with a PIM op. We refer to this model as the \textit{atomic model}. 

\textbf{Store model.} Another possible view of PIM ops in terms of existing memory operations can be as memory stores~\cite{PIMDB}. This follows the intuition that a PIM op does not read any data to the host processor. The PIM op only atomically writes to many addresses. Hence, PIM ops should have the same ordering rules as the host has for store operations. This view is more relaxed than the atomic model since store operations usually allow some reordering with other memory operations~\cite{PrimerBook,X86TSO}. One important difference between PIM and store operations is the address range they affect. Stores usually affect several bytes while PIM operations may affect an entire scope (up to the entire memory~\cite{CONCEPT,AMBIT}). As programs expect to read the last value written by the program order, memory operations to overlapping address ranges are not allowed to reorder. Hence, PIM ops must not reorder with memory operations to the same scope. We refer to this model as the \textit{store model}.

\textbf{Scope model.} Both atomic and store models assign familiar ordering rules to PIM ops, which might make PIM ops more intuitive to use as they will behave similarly to known operations. These ordering rules, however, ignore the unique characteristics of PIM ops. 
Treating PIM ops as their own class might produce a model that performs better and is better suited for PIM. We note that bulk-bitwise PIM ops are usually used because of their high throughput~\cite{SIMDRAM,AMBIT,FloatPIM,RACER,PIMDB}, performing numerous instances of the same computation concurrently, where the exact order of all the instances does not matter. These computations are performed on large data sets spanning many scopes, allowing the different scopes to process in any order. Hence, for the third consistency model, we allow PIM ops to reorder with any memory operation that is not assigned to their scope. Nevertheless, as with the store model, PIM ops are strictly ordered with operations to the same scope. To enforce the PIM ops order with other memory operations to other scopes, a dedicated fence should be used~\cite{OrderLight}. We refer to this model as the \textit{scope model}.

\textbf{Scope-relaxed model.} For the fourth model, we note that PIM ops usually do not operate on all the addresses within their scope~\cite{AMBIT,SIMDRAM,PIMDB,CONCEPT}, and it might be safe to reorder some PIM ops with other memory operations to the same scope. This knowledge -- which PIM ops and memory operations are safe to reorder -- is available to the software initiating the PIM ops. Hence, we take an example from existing relaxed consistency models~\cite{PrimerBook} and allow the software to indicate whether or not PIM ops and memory operations of the same scope can be reordered. In this model, PIM ops can be reordered with memory operations from other scopes and the same scope. To allow the software to enforce order between PIM ops and memory operations to the same scope, we include a new fence, termed the \textit{scope-fence}, which guarantees order only within a single scope. To enforce order between PIM ops and operations from other scopes, a dedicated fence (as in the scope model) is used. We refer to this model as the \textit{scope-relaxed model}.

To summarize, we propose four possible consistency models for PIM ops (see also Table~\ref{table:models}):

\begin{compactitem}
    \item \underline{Atomic model:} PIM ops are not allowed to reorder with any memory operation.
    \item \underline{Store model:} PIM ops take the same ordering rules as store operations in the host's consistency model.
    \item \underline{Scope model:} PIM ops can be reordered with any operations to other scopes, but not to the same scope.
    \item \underline{Scope-relaxed model:} PIM ops can be reordered with operations to other scopes and the same scope. Order is enforced by dedicated fence instructions.
\end{compactitem}

\section{Supporting Coherency}
\label{sec:coherency}

\begin{figure}[!t]
\centering
\includegraphics[width=\columnwidth]{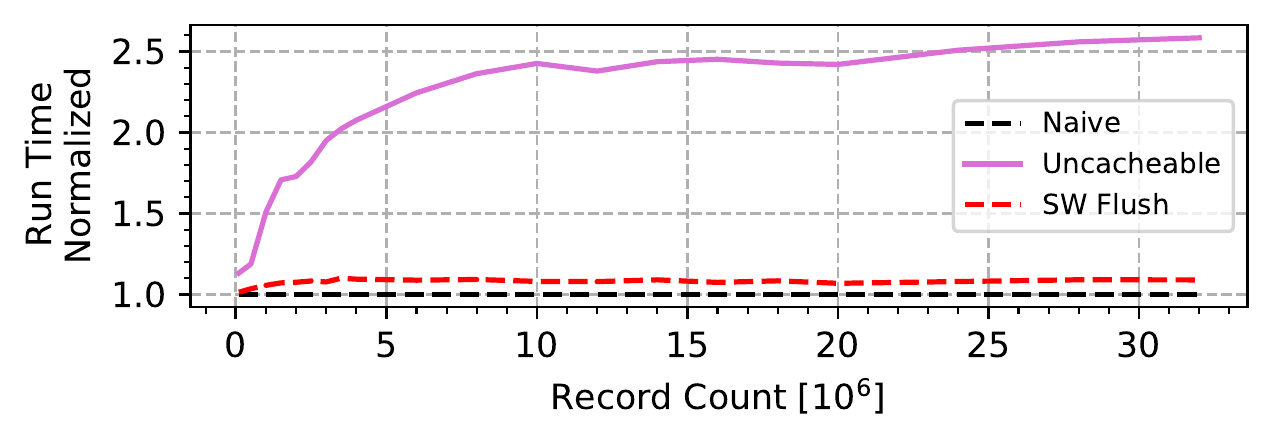}%
\caption{Run time of the YCSB workload on the Naive baseline (no coherency solution), the uncacheable approach (the Naive baseline where PIM-enabled scopes are uncacheable), and the Software Flush approach (flushing cache-lines from PIM-enabled scopes). The system and YCSB workload are detailed in Section~\ref{sec:methodology}.}
\label{fig:uc_compare}
\end{figure}

As identified in Section~\ref{sec:Intro}, to support any ordering rules with PIM ops, the cache flushes of affected addresses must be atomic with the PIM ops. The most straightforward solution is to mark PIM memory regions as uncacheable, as suggested for PIM techniques other than bulk-bitwise PIM~\cite{TRiM,RecNMP}. For bulk-bitwise PIM, however, reading the PIM results using memory loads is the most time-consuming step in the PIM computation~\cite{PIMDB}. Because these loads utilize spatial locality in the cache, making the PIM-operated memory region uncacheable will degrade performance substantially for bulk-bitwise PIM. 
Fig.~\ref{fig:uc_compare} compares the uncacheable approach and the software flush approach (described in Section~\ref{sec:Intro}) to a naive approach without any coherency solution (\emph{i.e.}, issuing PIM ops without a correctness guarantee); for details, see Section~\ref{sec:methodology}. As the PIM data set size increases, the PIM result size increases, requiring more data to be read from the PIM. Hence, the uncacheable approach run time becomes $2.57\times$ worse than the naive approach, while the software flush approach run time is only $1.09 \times$ worse.  

Rather than making the PIM-operated memory region uncacheable, we offer to support PIM coherency by enabling the PIM ops, on their way to the main memory, to flush all addresses from their scope. The scope flush might include data that is not used by the PIM op, an overhead of the scope abstraction (Section~\ref{sec:pim_consistency}). As the handling of PIM ops is a small part of the total execution time~\cite{PIMDB}, coupling the cache flushes with PIM ops is more promising than coupling them with the loads and stores (\emph{i.e.}, making data uncacheable).

The challenge in supporting cache flushes by PIM ops is the need to identify all cache-lines belonging to the PIM op scope. This is a challenge since scopes can be large, scattering cache-lines in all cache sets. Additionally, the cache structure does not inherently support a search for memory regions bigger than a cache-line. Thus, each PIM op is required to scan all cache sets one by one, and find the relevant cache-lines. This scan may take thousands of clock cycles, blocking the cache for the entire scan period and delaying other memory operations. To alleviate the cache scan overhead, two hardware techniques are suggested, the \textit{scope buffer} and the \textit{scope bit-vector}.

\subsection{Scope Buffer}
\label{subsec:scope_buffer}
\begin{figure}[!t]
\centering
\hspace{-8pt}
\begin{minipage}[c]{0.05\columnwidth}
\begin{subfigure}[c]{\textwidth}
\caption{}\label{fig:scopeCache_hitzero}
\end{subfigure}
\end{minipage}%
\begin{minipage}[c]{0.85\columnwidth}
\includegraphics[width=\textwidth,trim=0 0pt 0 4pt,clip]{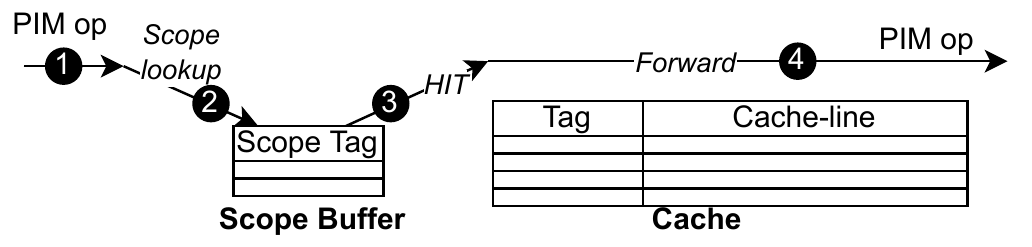}
\end{minipage}
\begin{minipage}[c]{0.05\columnwidth}
\begin{subfigure}[c]{\textwidth}
\caption{}\label{fig:scopeCache_miss}
\end{subfigure}
\end{minipage}%
\begin{minipage}[c]{0.85\columnwidth}
\includegraphics[width=\textwidth,trim=0 0pt 0 0pt,clip]{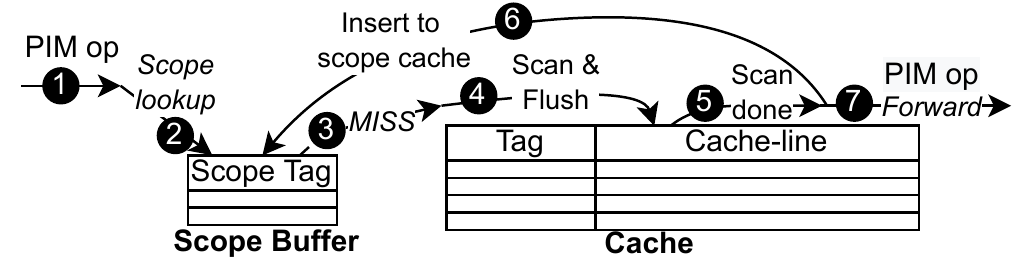}
\end{minipage}
\caption{Operation of PIM ops arriving at a cache with a scope buffer. Numbers in dark circles indicate the event sequence. (a) Hit in the scope buffer. 
(b) Miss in the scope buffer.}
\label{fig:scopeCache}
\end{figure}

We note that usually bulk-bitwise PIM has fine-grained instruction sets (\textit{e.g.}, AND, OR, NOT, ADD, MUL)~\cite{AMBIT,PIMDB,CONCEPT,SIMDRAM,Pinatubo}, requiring multiple PIM ops to perform a full computation. This suggests that if a PIM op for a scope is issued, then other PIM ops to the same scope will most likely follow, without intermediate loads or stores to the scope. This implies that PIM op scopes exhibit temporal locality and a single cache scan and flush can be performed for several consecutive PIM ops. To take advantage of this temporal locality, we propose adding a structure called the \textit{scope buffer} to the cache. The scope buffer's structure is similar to that of a cache~\cite{HWSW_IF} indexed by scope addresses and holding entries for scopes that were recently flushed from the cache. When a PIM op arrives at the cache, a lookup for its scope in the scope buffer is made. If the scope is found (Fig.~\ref{fig:scopeCache_hitzero}\ballnumber{1}\ballnumber{2}\ballnumber{3}), the PIM op is forwarded towards the main memory without performing a cache scan (Fig.~\ref{fig:scopeCache_hitzero}\ballnumber{4}). If the scope is not found in the scope buffer
(Fig.~\ref{fig:scopeCache_miss}\ballnumber{1}\ballnumber{2}\ballnumber{3}), a cache scan must be performed set-by-set, flushing all cache-lines of  that scope (Fig.~\ref{fig:scopeCache_miss}\ballnumber{4}). After the scan is finished (Fig.~\ref{fig:scopeCache_miss}\ballnumber{5}), the scope is inserted into the scope buffer (Fig.~\ref{fig:scopeCache_miss}\ballnumber{6}) and the PIM op is forwarded (Fig.~\ref{fig:scopeCache_miss}\ballnumber{7}). If the scope buffer is full and a new scope needs to be inserted into it, the new scope simply overwrites an old scope according to a replacement policy (\textit{e.g.}, LRU) with no additional action. Thus, when a stream of PIM ops to a single scope is sent to the memory, the first PIM op will perform a full cache scan and flush on the scope, and the rest of the PIM ops will pass through the cache without triggering further action.  

When a cache-line is inserted into the cache, a lookup request for the scope of this cache-line is made in the scope buffer. If found in the scope buffer, this scope is erased from the scope buffer with no additional action. If the scope is not found, nothing is done. Note that the interaction with the scope buffer for loads and stores takes place in parallel to the operations in the cache and off the critical path; thus, the cache operation latency is not affected.

To determine the required scope buffer size, we noted the following. First, after each PIM computation the results are read from all involved scopes, invalidating these scopes from the scope buffer before the next PIM computation. Second, during a single PIM computation, a scope in the scope buffer is useful only during the issuing of PIM ops to that scope. Therefore, after PIM ops to a scope are done issuing, the scope entry in the scope buffer is not useful anymore.
Hence, the scope buffer is only required to hold the scopes that are currently being issued with PIM ops. The number of such scopes depends on the processor configuration and software code. Similarly to choosing a cache size~\cite{QuantApproch}, the scope buffer size can be set according to the processor configuration. We have seen that even a small-sized scope buffer is sufficient to achieve close to the maximum possible hit rate for all data set sizes (see Section~\ref{sec:eval}).    

\subsection{Scope bit-vector (SBV)}
\label{subsec:scope_bit_vector}
Although the scope buffer reduces the number of cache scans, the remaining cache scans' latency can reach thousands of clock cycles. We note that for bulk-bitwise PIM, the results of a computation are stored on multiple crossbar arrays at the same specific crossbar columns and rows~\cite{SIMDRAM,PIMDB,RACER}, giving a regular non-continuous address range for the result because of the address mapping of crossbar columns and rows~\cite{SIMDRAM,PIMDB}. The consequence of PIM results having regular non-continuous address ranges is that a PIM result tends to cluster in a subset of cache sets.
The cache sets in this subset depend on the workload and architecture, potentially including any cache set, but tending to be less than all cache sets.

To take advantage of the fact that cache-lines from PIM-enabled scopes tend to cluster in a subset of the cache sets, we propose an additional structure that keeps track of these cache sets. We name this structure the \textit{scope bit-vector} (SBV). As the name suggests, the SBV is a bit-vector with a single bit for each cache set. A bit in the SBV is set high if its corresponding cache set contains a cache-line from some PIM-enabled scope; otherwise, the bit is kept low. Using the SBV, a cache scan for a PIM op is required to check only the sets that have a high bit in the SBV. To enable the use of the SBV, a cache-line from a PIM-enabled scope must be marked as such. This marking can be implemented, for example, by defining a memory page as PIM-enabled, encoding this information in the translation table page entry, attaching it to every memory request for this page (similar to marking a page as uncacheable~\cite{QuantApproch}), and adding this information to each cache-line's meta-data. In this way, when a cache-line from a PIM-enabled scope is inserted into the cache, the relevant bit in the SBV is turned to high. When a cache-line from a PIM-enabled scope is evicted, all remaining cache-lines in the same set are checked to see whether at least one belongs to some PIM-enabled scope and the SBV bit is updated accordingly.

To summarize the above, to maintain coherency between the host caches and the PIM module, we suggest that PIM ops flush cache-lines from their scope on their way to the memory. When a PIM op arrives at the cache, a lookup for the scope of the PIM op is made in the scope buffer. The scope buffer lookup answer indicates if the PIM op can be forwarded or if a cache scan is required to find and flush all cache-lines from the scope of the PIM op. If a scan is required, the SBV is used to identify the cache sets to scan. 

\section{Supporting PIM Consistency Models}
\label{sec:consistency_support}
Supporting coherence is not enough to enforce the consistency models suggested in Section~\ref{sec:pim_consistency}.
How the host cores, memory subsystem, and the PIM module handle PIM ops must also be specified.
Therefore, in this section, we present an implementation for each consistency model. These implementations require only a minimal hardware overhead while still enabling the reordering allowed by each consistency model. Table~\ref{table:models} summarizes the implementation for each model.

\subsection{Base Host Processor and PIM Module}
\label{subsec:host_assumptions}
\begin{figure}[!t]
\centering
\includegraphics[width=0.7\columnwidth]{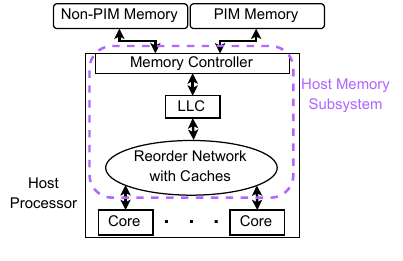}%
\caption{The host and system used to demonstrate the proposed bulk-bitwise PIM consistency models.}
\label{fig:system_model}
\end{figure}

For the host of the PIM module, we take a multicore system illustrated in Fig.~\ref{fig:system_model}, as is commonly used for bulk-bitwise PIM architectures~\cite{CONCEPT,PIMDB,SIMDRAM}. Other hosts (\textit{e.g.}, GPU~\cite{OrderLight}) can be used similarly. To keep our implementation general and not rely on host-specific attributes, we assume the host has the following common capabilities: (1) The host instruction set contains a dedicated instruction to issue a PIM op (similarly to a store instruction issuing a memory write)~\cite{PIMDB,AMBIT,SIMDRAM,OrderLight,Ahn2015,LazyPIM}. (2) Host cores can execute out-of-order and issue non-PIM memory operations according to the host's native (non-PIM) consistency model~\cite{PrimerBook}. (3) The host memory subsystem can reorder operations passing through it, \textit{e.g.}, by a multi-path network-on-chip~\cite{NoCbook}, virtual channels~\cite{NoCbook}, or non-FIFO buffers~\cite{PrimerBook}. (4) The host memory subsystem consists of an inclusive, shared last-level-cache (LLC), as is prevalent in modern processors~\cite{QuantApproch}. The rest of the cache levels are possibly private or shared. (5) The host memory controller can reorder operations, but does not violate data dependencies between operations. For example, two writes to the same address will be executed in the order of arrival to the memory controller, but writes to different addresses can be reordered. Similarly, the memory controller will not reorder a PIM op with other memory operations that address the same scope.

As the PIM module itself may contain additional routing and logic to handle memory operations~\cite{PIMDB,CONCEPT}, it is assumed that the PIM module, like the host's memory controller, does not violate data dependencies between operations.
Hence, when a PIM op arrives at the host memory controller, it cannot reorder with other memory operations. To understand why, consider the scenario where the memory controller received PIM op $A$, and had not yet forwarded it when another memory operation $B$ is issued from another thread, not necessarily arriving at the memory controller. To see if $B$ happened before $A$, a thread needs to see that $B$ happened and $A$ has not. To do this, another memory operation $C$ has to be sent to $A$'s scope (\emph{e.g.}, a read) after $B$; however, then $C$ will arrive at the memory controller after $A$, and as $A$ and $C$ operate on the same scope, the memory controller and the PIM module will keep their order, making $C$ operate after $A$, and see its results. Consequently, no thread can see $B$ happening before $A$. This means that when a PIM op arrives at the memory controller, it is safe for the issuing thread to continue issuing other memory operations without the risk of violating memory ordering with that PIM op.
Therefore, to enforce the required order of PIM ops and other memory operations, it is sufficient to enforce it only between the host cores and the memory controller, \textit{i.e.}, enforcing the order in the cores, caches, and the on-chip network. 

\subsection{Atomic Model}
To support the atomic model with minimal modifications to the host, we add a scope buffer and an SBV only to the LLC. Since the LLC is inclusive, flushing cache-lines from the LLC will automatically flush the cache-lines from all cache levels, saving the need to flush each cache level in turn. Additionally, the LLC is the largest cache level, making the added hardware and complexity overheads more reasonable. 

\begin{figure}[!t]
\centering
\subfloat[\label{fig:atomic_model}]{\includegraphics[trim=0 3pt 0pt 0pt, clip,width=0.25\columnwidth]{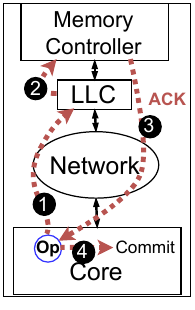}}%
\subfloat[\label{fig:write_scope_model}]{\includegraphics[trim=0 3pt 0pt 0pt, clip,width=0.25\columnwidth]{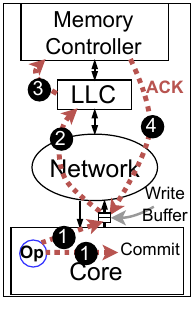}}%
\subfloat[\label{fig:scope_relax_op}]{\includegraphics[trim=0 3pt 0pt 0pt, clip,width=0.25\columnwidth]{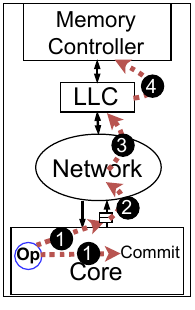}}%
\subfloat[\label{fig:scope_realx_fence}]{\includegraphics[trim=0 3pt 0pt 0pt, clip,width=0.25\columnwidth]{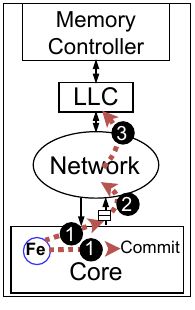}}
\caption{Consistency model implementations. \protect\ballblue{Op} indicates a PIM op. \protect\ballblue{Fe} indicates a scope-fence. (a) Atomic model's PIM op issuing process.
(b) Store and Scope models' PIM op issuing process.
(c) Scope-relaxed model's PIM op issuing process.
(d) Scope-relaxed model's scope-fence issue process. See Section~\ref{sec:consistency_support} for details.
}
\label{fig:models}
\end{figure}

The process of issuing a PIM op in the atomic model is shown in Fig.~\ref{fig:atomic_model}. Here, a core treats a PIM op instruction as a memory operation with a memory fence before and after it, not allowing any memory operation to reorder with the PIM op. When it is safe to commit the PIM op, it is issued to the host's memory subsystem (Fig.~\ref{fig:atomic_model}\ballnumber{1}). The core, however, does not yet commit the PIM op, thereby keeping the thread from issuing other memory operations and preventing their reordering with the PIM op at the memory subsystem. The PIM op is forwarded to the LLC without affecting the lower cache levels. When the PIM op arrives at the LLC, the LLC flushes all cache-lines from the scope of the PIM op to the memory as explained in Section~\ref{sec:coherency}. When the flush is complete, the LLC forwards the PIM op to the memory controller (Fig.~\ref{fig:atomic_model}\ballnumber{2}). Once the PIM op arrives at the memory controller and ensures the order of the PIM op, the memory controller sends an ACK to the issuing core (Fig.~\ref{fig:atomic_model}\ballnumber{3}). Upon receiving the ACK, the core commits the PIM op instruction (Fig.~\ref{fig:atomic_model}\ballnumber{4}).

\subsection{Store Model}
The store model is supported similarly to the atomic model. The difference is that a core treats a PIM op as if it were a store operation, possibly allowing reordering with operations outside its scope (depending on the host's reordering rules for stores). Additionally, between the time a PIM op is issued to the memory subsystem and the time its ACK is received, the core can still issue operations that are allowed to bypass PIM ops. For example, in an X86-TSO core~\cite{X86TSO}, a load can bypass a store to a different address. Therefore, a load to a scope different than the PIM op's scope can still be issued to memory while the PIM op instruction waits for its ACK.   

Fig.~\ref{fig:write_scope_model} shows the process of the store model for PIM op issuing. When it is safe for the core to commit a PIM op, it is both issued to the host's memory subsystem and committed (Fig.~\ref{fig:write_scope_model}\ballnumber{1}). To prevent reordering by the memory subsystem, the entry point to the memory subsystem (\textit{e.g.}, the write buffer~\cite{PrimerBook,QuantApproch}) enforces the ordering rules. After sending a PIM op (Fig.~\ref{fig:write_scope_model}\ballnumber{2}), the entry point must prevent relevant memory operations from entering the memory subsystem. 

Similarly to the atomic model, a PIM op flushes only the LLC and is forwarded to the memory controller afterwords (Fig.~\ref{fig:write_scope_model}\ballnumber{3}). The memory controller then sends an ACK to the memory subsystem entry point  (Fig.~\ref{fig:write_scope_model}\ballnumber{4}), indicating that it is safe to issue the withheld memory operations.
This process of PIM op issuing is similar to a store operation's issuing process with a write buffer~\cite{PrimerBook} -- only in our case, the host memory controller, rather than the L1 cache, indicates completion. 

\subsection{Scope Model}
The scope model is supported similarly to the store model and as shown in Fig.~\ref{fig:write_scope_model}. The difference is that we want to allow reordering for PIM ops with memory operations to other scopes. Hence, the memory subsystem entry point is required to hold back only operations to a scope of an ongoing PIM op. Other operations can enter the memory subsystem without waiting for the PIM op's ACK (\textit{i.e.}, a non-FIFO write buffer). To enforce ordering between PIM ops of different scopes, the dedicated fence from~\cite{OrderLight} is used. By allowing more memory operations to clear the memory subsystem entry point, the entry point can receive more operations from the core, preventing the core from stalling on resource contention. Allowing the memory subsystem to handle more operations concurrently, while the core continues execution, potentially leads to better system utilization and performance.

\subsection{Scope-Relaxed Model}
The scope-relaxed model allows PIM ops to reorder with memory operations to the same scope, hence cores can reorder PIM ops with all other memory operations. Thus, the scope-relaxed model's process of PIM op issuing, shown in Fig.~\ref{fig:scope_relax_op}, allows the core to issue a PIM op at commit (Fig.~\ref{fig:scope_relax_op}\ballnumber{1}), and also does not require the host's memory subsystem entry point to hold back any memory operation when forwarding a PIM op (Fig.~\ref{fig:scope_relax_op}\ballnumber{2}). This enables multiple PIM ops, from the same and different scopes, to be inserted into the memory subsystem at the same time and cleared from the entry point, increasing system utilization compared to the previous models. As with the previous models, to preserve the atomicity of PIM ops and cache flushes, PIM ops must flush their scope from the LLC (Fig.~\ref{fig:scope_relax_op}\ballnumber{3}) before being forwarded to the memory controller (Fig.~\ref{fig:scope_relax_op}\ballnumber{4}). Unlike the previous models, however, PIM ops must pass through all cache levels on their way to the LLC (Fig.~\ref{fig:scope_relax_op}\ballnumber{2}) without flushing them (reason given below) and do not require the memory controller to return an ACK.

\begin{table}[t]
\setlength\tabcolsep{3pt}
\centering
\begin{tabular}{|c|G|H|J|} 
\hline
\rule{0pt}{1.8ex} \textbf{Model}  & \textbf{PIM Op Allowed Reordering} & \textbf{Additional Fence Required} & \textbf{Scope Buffer \& SBV}\\ \hline
\rule{0pt}{1.8ex} Atomic & None & No &  Only LLC \\\hline
\rule{0pt}{1.8ex} Store & Same as store operations & No & Only LLC \\\hline
\rule{0pt}{1.8ex} Scope & All operations to other scopes & Ordering between scopes & Only LLC \\ \hline
\rule{0pt}{1.8ex} Scope-Relaxed & All operations except fences  & (1) Ordering within scope and (2) between scopes & All caches\\ \hline
\end{tabular}
\caption{Consistency model definitions and implementations.
}
\label{table:models}
\end{table}
\setlength\tabcolsep{6pt}

To enforce ordering between PIM ops and memory operations from the same scope, a scope-fence, implemented as in~\cite{OrderLight}, is used (Section~\ref{sec:pim_consistency}). The scope-fence's issuing process is similar to that of a PIM op's, shown in Fig.~\ref{fig:scope_realx_fence}. The scope-fence is issued from the core at commit time (Fig.~\ref{fig:scope_realx_fence}\ballnumber{1}). Also, the scope fence passes through all cache levels on its way to the LLC (Fig.~\ref{fig:scope_realx_fence}\ballnumber{2}). Unlike a PIM op, however, the scope-fence propagates through all optional paths to a destination~\cite{OrderLight}, \textit{i.e.,} the next cache level, duplicating the scope-fence packet as necessary. On the way to a destination, all memory operations to the scope-fence's scope are not allowed to reorder around the scope-fence in any path. Once all copies arrive at the destination, the scope-fence is forwarded to the next destination (through all optional paths). The scope-fence is terminated at the LLC (Fig.~\ref{fig:scope_realx_fence}\ballnumber{3}).  Also, unlike a PIM op, a scope-fence must flush its scope in all caches on its path. Otherwise, the following may happen. 
Consider a scenario where a PIM op, a scope-fence, and a load, all from the same thread and to the same scope, are issued in that order from a core. Since the issued PIM op and scope-fence do not block the load in the scope-relaxed model, the load can be issued. The load may hit in a lower level cache (skipped by the scope-fence) before the PIM op has reached the LLC and flushed the load's data from the caches, view the pre-PIM value, and effectively be ordered before the PIM op. This, of course, breaks the ordering guarantee of the scope-fence.

Hence, all caches in this implementation include a scope buffer and an SBV. PIM ops perform a cache scan only at the LLC, while scope-fences perform a cache scan in all caches on the path between the issuing core and the memory controller. Note that without a scope-fence, the scope-relaxed model allows the load in the previous scenario to reorder with the PIM op, so the PIM op is not required to perform a scan on all caches. PIM ops, however, do need to be routed through all cache levels on their way to the LLC, so they might be ordered by a scope-fence. 
To enforce order between PIM ops from different scopes, an additional fence with the same implementation of~\cite{OrderLight} is used (as in the scope model), affecting all memory operations from all scopes.

\begin{table}[t]
\setlength{\tabcolsep}{2pt}
\small
\centering
\begin{tabular}{|A|B||E|F|} 
\hline
\multicolumn{4}{|c|}{\rule{0pt}{1.6ex}Evaluation System}\\
\hline 
\rule{0pt}{1.8ex}Processor Cores& \mbox{6 cores, X86,} \mbox{OoO, 3.6GHz}  & Main Memory  & \mbox{32GB DRAM}, \mbox{DDR4-2400}  \\\hline
\rule{0pt}{1.8ex}   L1 cache & \mbox{Private, 16KB,} \mbox{64B block}, 4-way  & L2 cache & \mbox{Shared, 2MB,} \mbox{64B block}, 16-way \\\hline
\rule{0pt}{1.8ex}   L1 scope buffer \mbox{(if exists)}  & \mbox{16 sets,} \mbox{1-way}& L2 scope buffer & \mbox{64 sets, 4-way} \\\hline
\rule{0pt}{1.8ex} Coherency protocol  & MESI  & PIM modules & 1 \mbox{(spec as in~\cite{PIMDB})}\\\hline
\rule{0pt}{1.8ex} Scope  & 2MB huge page & Max. database records per scope & 32K\\\hline

\end{tabular}
\caption{Architecture and system configuration to capture the consistency model behavior.}
\label{table:arch_details}
\end{table}

\section{Methodology}
\label{sec:methodology}
\subsection{System Simulation}
To compare the performance and behavior of our consistency models and coherence solutions, we developed a bulk-bitwise PIM simulator\footnote{\mbox{Available at:}\newline\hspace*{1.2em} \mbox{\url{https://github.com/benperach/gem5_bulkbitwise_PIM_consistency}.}} based on the gem5 simulator~\cite{gem5}. We adopted the solution of~\cite{PIMDB} as our PIM module, since, to the best of our knowledge, this is the only available bulk-bitwise PIM solution using virtual memory. All consistency models were implemented as described in Section~\ref{sec:consistency_support} with a six-core multicore and a two-level cache hierarchy (L2 is the LLC) with a MESI coherency protocol~\cite{PrimerBook}. All simulations were performed on the gem5 full-system mode, running a Linux kernel, and using the gem5 Ruby memory system. The details of the simulated system are listed in Table~\ref{table:arch_details}. Estimating the scope-buffer and SBV hardware overhead for the L2 cache, using the Synopsys 28nm library, we observe a $0.092\%$ area overhead. The scope-relaxed model requires a scope-buffer and an SBV for all caches, reaching a total of $0.22\%$ area overhead.

\subsection{Workload}
\label{subsec:workload}
Previous works suggest that database workloads, specifically scanning operations, are a promising application for bulk-bitwise PIM~\cite{AMBIT,CONCEPT,SIMDRAM,PIMDB}. Hence, as a representative workload for bulk-bitwise PIM, we take the YCSB short-range workload~\cite{YCSB}, a key-value database workload divided into $95\%$ scan operations and $5\%$ record insertions, and queries from the TPC-H~\cite{TPCH}, an online analytical database benchmark. Other workloads from the YCSB benchmark were not used as they do not involve bulk-bitwise PIM operations.

\begin{table}[t]
\setlength\tabcolsep{3pt}
\centering
\begin{tabular}{|c|c|} 
\hline
\rule{0pt}{1.8ex} Number of Operations    & 1000    \\ \hline
\rule{0pt}{1.8ex} Scan Operation Percentage & 95\%    \\\hline
\rule{0pt}{1.8ex} Insert Operation Percentage & 5\%	\\\hline
\rule{0pt}{1.8ex} Number of Fields per Record  & 5 \\ \hline
\rule{0pt}{1.8ex} Field Length & 10B    \\ \hline
\rule{0pt}{1.8ex} Records in Scan Results & Uniform dist. [1,100]   	\\ \hline
\rule{0pt}{1.8ex} Scan Base Record & Zipfian dist.   	\\ \hline
\end{tabular}
\caption{YCSB~\cite{YCSB} workload summary. The number of records varies with experiments.}
\label{table:YCSB}
\end{table}
\setlength\tabcolsep{6pt}

\begin{table}[t]
\setlength\tabcolsep{3pt}
\centering
\begin{tabular}{|c|c|c||c|c|c|} 
\hline
\rule{0pt}{1.8ex} \textbf{Query}  & \textbf{\# Scopes} & \textbf{PIM section} & \textbf{Query}  & \textbf{\# Scopes} & \textbf{PIM section}  \\ \hline
\rule{0pt}{1.8ex} q1 & 1832 &  Full-query & q12 & 1832 & Filter only\\\hline
\rule{0pt}{1.8ex} q2 & 66 &	Filter only & q14 & 1832 & Filter only\\\hline
\rule{0pt}{1.8ex} q3 & 2336 & Filter only & q15 & 1832 & Filter only\\ \hline
\rule{0pt}{1.8ex} q4 & 2290 & Filter only & q16 & 62 & Filter only\\ \hline
\rule{0pt}{1.8ex} q5 & 508  & Filter only & q17 & 62 & Filter only\\ \hline
\rule{0pt}{1.8ex} q6 & 1832 & Full-query & q19 & 1894 & Filter only\\ \hline
\rule{0pt}{1.8ex} q7 & 1882 & Filter only & q20 & 2294 & Filter only\\ \hline
\rule{0pt}{1.8ex} q8 & 566 & Filter only & 21 & 1832 & Filter only\\ \hline
\rule{0pt}{1.8ex} q10 & 2290 & Filter only & q22 & 46 & Full sub-query\\ \hline
\rule{0pt}{1.8ex} q11 & 4 & Filter only & & &\\ \hline
\end{tabular}
\caption{TPC-H~\cite{TPCH} query summary. Queries 9, 13, and 18 do not have a PIM section~\cite{PIMDB} and are, therefore, not evaluated.
}
\label{table:TPCH}
\end{table}
\setlength\tabcolsep{6pt}

In the YCSB benchmark, a database scan operation searches for a set of records from a single database relation (organized as in~\cite{PIMDB}) and extracts a certain text field from each found record. For each database scan, the number of result records and the field to extract are randomly generated. 
To measure this workload on the suggested consistency models, we used runs of 1000 operations ($95\%$ scans, $5\%$ insertions, randomly ordered) on a varying number of records. All database scans, reads, and insertions were performed within the PIM module memory. The parameters of the workloads are summarized in Table~\ref{table:YCSB}.
Insertions were performed using standard stores, and scans were performed by: (1) Dividing the scopes of the database evenly among four threads, (2) having each thread issue PIM ops to perform the scan on each of its assigned scopes, 
and (3) having each thread read the scan result and the required record fields from the database contained in its assigned scopes using standard loads.

To show the performance and behavior trends of our four models, we ran the YCSB workload on a range of database sizes, from $0.1\times10^6$ to $32\times10^6$ records, occupying from $4$ to $977$ scopes. For all scope counts and all models, the same sequence of scans and insertions was measured. Records are randomly distributed in the database, making the scan result evenly distributed across the scopes. The range of scope counts was chosen to capture the behavior of our models. As shown in Section~\ref{sec:eval}, the measurements reached a steady trend on the high end of the scope count range, allowing us to focus on the above-mentioned scope count range.

For the TPC-H queries, each query was run ten times consecutively. Each query run was performed as in~\cite{PIMDB}, executing only the PIM section of the query and reading the results. The PIM section of the query is either only filtering the involved database relations or performing the entire query when a single relation is involved. Queries 9, 13, and 18 were not performed since they do not include any PIM section. Table~\ref{table:TPCH} lists the TPC-H queries key parameter; see~\cite{PIMDB} for details.

An important difference between the TPC-H and YCSB workloads is in the number of reads performed after each PIM computation. For the TPC-H, only the PIM computation result is read, resulting in a regular read pattern that is mapped to a limited group of cache sets (see Section~\ref{subsec:scope_bit_vector}). For the YSCB workload, the result of the PIM computation indicates what other data from the PIM memory have to be read. The latter data have a different access pattern and are mapped to a different group of cache sets, creating more work for the PIM coherence mechanism (\textit{e.g.}, software flushes, cache scans).

\subsection{Comparison Baselines}
We compared the proposed consistency models and their implementations to two baselines. The first baseline is the software flush approach presented in Section~\ref{sec:Intro} and used in previous bulk-bitwise works~\cite{PIMDB,SIMDRAM}. In this approach, the software running on the host is responsible for maintaining coherency by explicitly issuing cache-line flushes. Host cores issue PIM ops to the memory subsystem on commit. Thereafter, PIM ops are forwarded directly to the memory controller without performing any operation in the memory subsystem. As discussed in Section~\ref{sec:Intro}, such an approach cannot guarantee correctness. We refer to this baseline as \textit{SW Flush}.

The second baseline is a naive approach where the cache is not flushed at all, \textit{i.e.}, the program is executed as in the SW Flush baseline, only without the software flushes. Cache-lines are evicted only by normal operation during loads and stores. This baseline does not provide correct execution; it serves to show the performance overhead of the different consistency models. We refer to this baseline as \textit{Naive}.
\section{Evaluation}
\label{sec:eval}

\begin{figure}[!t]
\centering

\begin{minipage}[c]{0.05\columnwidth}
\begin{subfigure}[c]{\textwidth}
\caption{}
\end{subfigure}
\end{minipage}\hfill%
\begin{minipage}[c]{0.88\columnwidth}
\includegraphics[width=\textwidth,trim=0 0pt 0 0pt,clip]{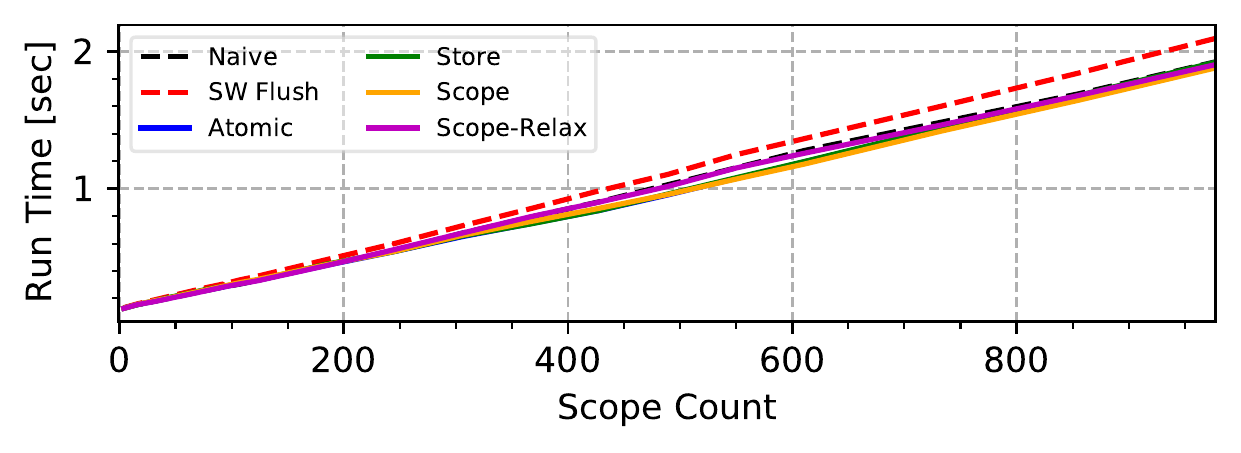}
\end{minipage}\\

\begin{minipage}[c]{0.05\columnwidth}
\begin{subfigure}[c]{\textwidth}
\caption{}\label{subfig:runtime_norm}
\end{subfigure}
\end{minipage}%
\begin{minipage}[c]{0.95\columnwidth}
\includegraphics[width=\textwidth,trim=0 0pt 0 0pt,clip]{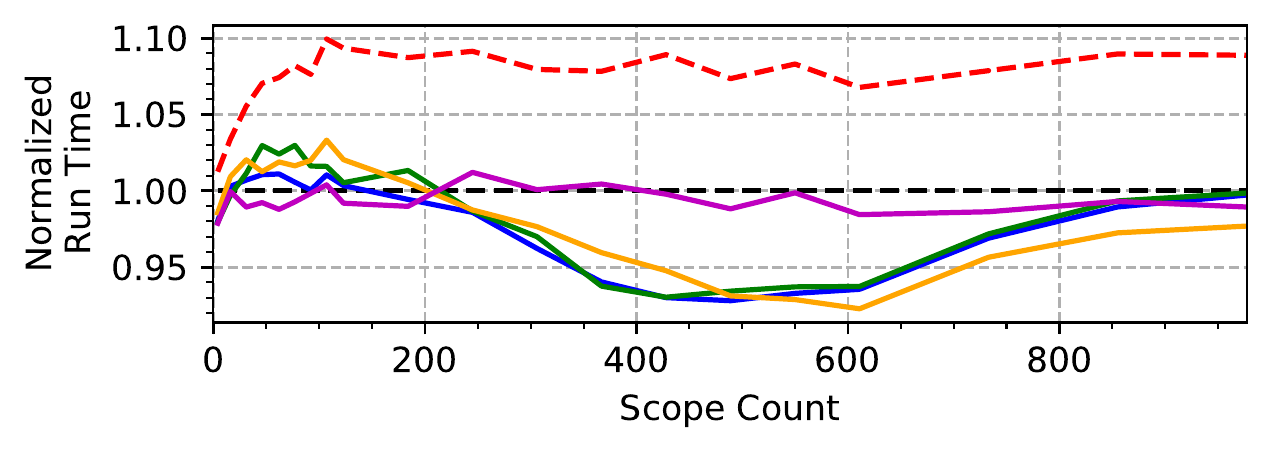}
\end{minipage}
\caption{YCSB benchmark. (a) Absolute run time. (b) Run time normalized to the Naive baseline. The Naive and SW Flush baselines (dashed lines) do not guarantee correct execution.}
\label{fig:timing_ratio}
\end{figure}

\begin{figure*}[!t]
\centering
\includegraphics[trim=0 0 0 7pt, clip,width=2\columnwidth]{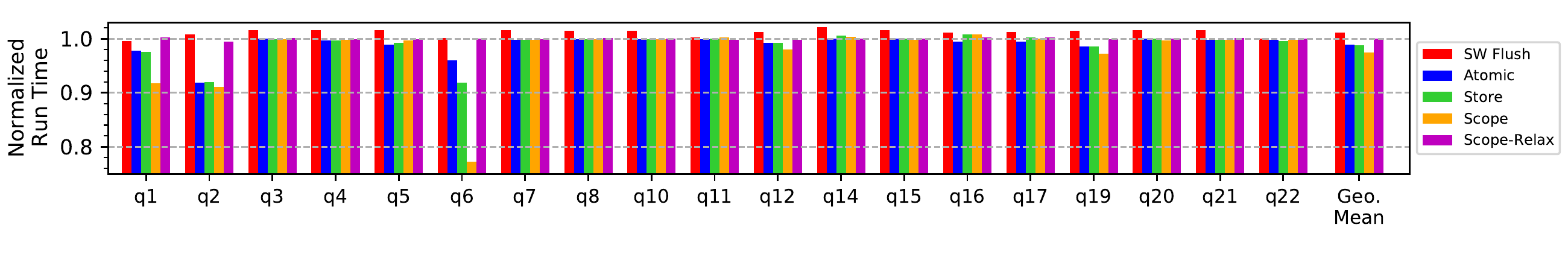}
\caption{Run time for TPC-H queries normalized to the Naive baseline. The Naive and SW Flush baselines do not guarantee correct execution.}
\label{fig:timing_ratio_tpch}
\end{figure*}

\begin{figure}[!t]
\centering
\includegraphics[trim=0 0 0 8pt, clip,width=\columnwidth]{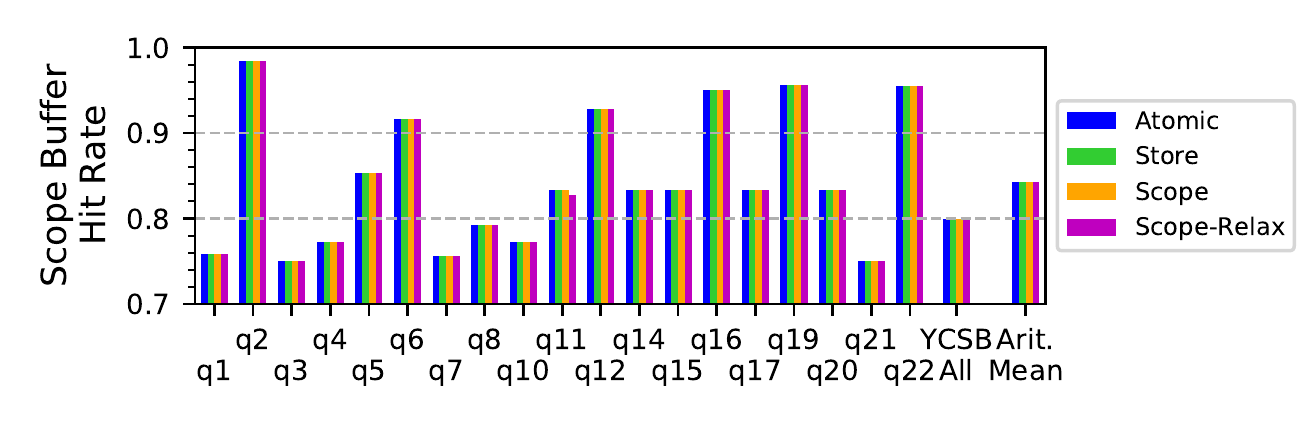}
\caption{Scope buffer hit rate for TPC-H and YCSB.}
\label{subfig:scope_hitrate}
\end{figure}

\begin{figure}[!t]
\centering
\begin{minipage}[c]{0.05\columnwidth}
\begin{subfigure}[c]{\textwidth}
\caption{}\label{subfig:buffer}
\end{subfigure}
\end{minipage}%
\begin{minipage}[c]{0.42\columnwidth}
\includegraphics[width=\textwidth,trim=0 4pt 0 8pt,clip]{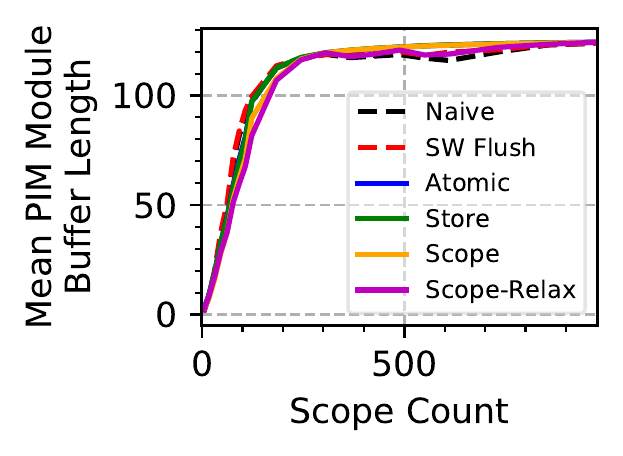}
\end{minipage}%
\begin{minipage}[c]{0.05\columnwidth}
\begin{subfigure}[c]{\textwidth}
\caption{}\label{subfig:buffer_scope}
\end{subfigure}
\end{minipage}%
\begin{minipage}[c]{0.42\columnwidth}
\includegraphics[width=\textwidth,trim=0 4pt 0 8pt,clip]{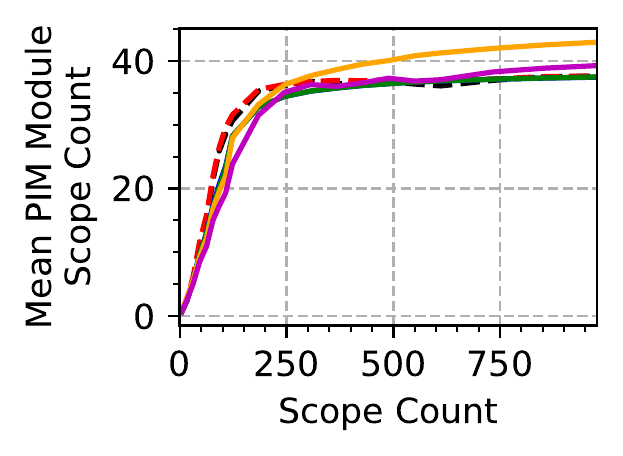}
\end{minipage}\\[4pt]

\begin{minipage}[c]{0.05\columnwidth}
\begin{subfigure}[c]{\textwidth}
\caption{}\label{subfig:Scan_cycles}
\end{subfigure}
\end{minipage}%
\begin{minipage}[c]{0.42\columnwidth}
\includegraphics[width=\textwidth,trim=0 4pt 0 6pt,clip]{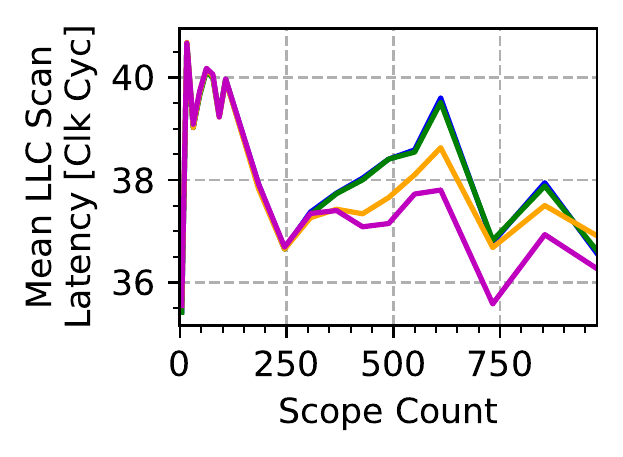}
\end{minipage}%
\begin{minipage}[c]{0.05\columnwidth}
\begin{subfigure}[c]{\textwidth}
\caption{}\label{subfig:sbv_skipratio}
\end{subfigure}
\end{minipage}%
\begin{minipage}[c]{0.42\columnwidth}
\includegraphics[width=\textwidth,trim=0 4pt 0 6pt,clip]{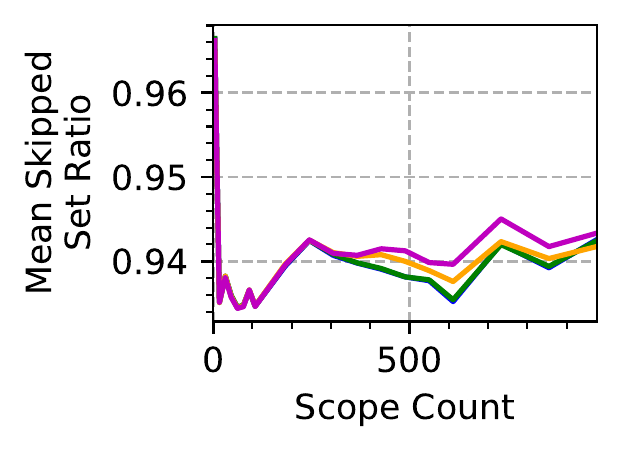}
\end{minipage}%

\caption{System statistics for the YCSB workload. (a) Mean PIM module buffer size on PIM op arrival. (b) Mean number of unique scopes at the PIM module buffer on PIM op arrival. (c) Mean LLC scan latency. PIM ops not requiring any scan (scope buffer hit) are counted as zero clock cycle scans. (d) SBV mean ratio of LLC sets skipped out of all LLC sets during an LLC scan.
}
\label{fig:LLC_scan}
\end{figure}

We start by comparing the four proposed models and the two baselines. Fig.~\ref{fig:timing_ratio} and Fig.~\ref{fig:timing_ratio_tpch} show the run times of the YCSB and TPC-H workloads, respectively, for all models and baselines. Fig.~\ref{subfig:scope_hitrate} and Fig.~\ref{fig:LLC_scan} show some system statistics.

The YCSB workload (Fig.~\ref{fig:timing_ratio}) shows  several regions of interest. In the first region, up to $100$ scopes, the run time of the four models and the SW Flush baseline increase relative to the Naive baseline. In this region, the number of used PIM-enabled scopes is relatively low, resulting in a lightly loaded system and a low number of load operations to memory to retrieve the PIM results. This makes the overhead of managing the PIM ops more prominent, as shown by the increasing performance benefits of the Naive baseline. 

For more than $100$ scopes, the relative overhead of the SW Flush is constant. As the number of flush operations is proportional to the workload size, the flush relative overhead becomes constant when the workload size-dependent execution (\textit{e.g}, PIM computation, result read) becomes the dominant part of the run time.

Additionally, from approximately $100$ scopes,
the performance of the four models improves relative to the Naive baseline.
The reason for this is that with the increasing scope count, more PIM ops are required. As PIM ops have a long PIM execution time, the host cores, in all models, issue PIM ops at a higher rate than the PIM module can process them. As a result, the PIM module buffer is filled (Fig.~\ref{subfig:buffer}), back-pressuring the host memory subsystem and saturating it.  
In the Naive and SW Flush baselines, cores issue PIM ops at a fast rate since there are no constraints on PIM ops. In the four consistency models, however, the ordering mechanisms throttle the cores' PIM ops issue rate, resulting in a lighter load on the host's memory subsystem and allowing other memory operations (\textit{e.g.,} reads from other threads) to execute quicker. The scope-relaxed model behaves similarly to the Naive baseline since it too allows a high PIM op issue rate, also saturating the memory subsystem.
We investigate the effect of the PIM module buffer further below.

For more than $600$ scopes, the improvements in the models' run time start to diminish. The benefit from lighter loads in the memory subsystem is relevant when one thread issues PIM ops while another thread issues read operations. The lighter load allow the threads to execute concurrently by interleaving their operations. This overlap, however, shortens, relative to the whole run time, as the workload size increases. Thus, when the workload size increases, the run time differences depend mostly on the execution time when all threads issue PIM ops. When all threads issue PIM ops, the scope model, due to its inherent interleaving of PIM ops from different scopes, has the best run time. In the scope model, the core's write-buffer holds back PIM ops that have an ongoing PIM op to their scope, but allows other PIM ops to continue. This results in PIM ops from different scopes arriving at the PIM module's buffer in an interleaved manner, shown in Fig.~\ref{subfig:buffer_scope} as the increased number of unique scopes in the PIM module's buffer. Having PIM ops to more scopes, the PIM module can concurrently execute more PIM ops, increasing parallelism and shortening total execution time. The other models issue PIM ops in program order, resulting in PIM ops from the same scope reaching the PIM module together and executing serially.

The relative run time of the models on the TPC-H queries is shown in Fig.~\ref{fig:timing_ratio_tpch}. The models show little run time difference on most queries. When the difference between the models is significant, the scope model has the best run time, followed by the store and atomic models; the scope-relaxed model is always similar to the Naive baseline. The same effects described above for the YCSB workload also apply to the TPC-H queries. Queries q1, q2, q6, q12, q14, q15, q19, and q20 are the only queries that reach a substantial PIM module buffer occupancy, due to a combination of the number of scopes, PIM ops per scope, and PIM ops latency (not shown). Queries q14, q15, and q20 have a few PIM ops per scope and a relatively short PIM execution time per scope, easing the back-pressure
created by the full buffer to the point that all models perform similarly. For q1, q2, q6, q12, and q19 we see a visible difference between the models. Queries q2, q12, and q19 have more and longer PIM ops per scope relative to other filter-only queries. Furthermore, q2 has few scopes, reducing the resulting read phase's execution time. For q1 and q6, being full-queries, the PIM section is substantially longer and there are fewer results to read~\cite{PIMDB}. For these reasons, queries q1, q2, q6, q12, and q19 have a more substantial PIM section and thus a performance difference between the models is visible.

\textbf{LLC Scan:} Fig.~\ref{subfig:scope_hitrate} and Fig.~\ref{fig:LLC_scan} shows statistics of LLC scans and the performance of the scope-buffer and SBV at the LLC. Since the scope buffer is large enough to hold all concurrently issued scopes in all models, the first PIM op to a scope misses in the scope buffer while all the other PIM ops to that scope hit. This results in the same hit rate for all models, as shown in Fig.~\ref{subfig:scope_hitrate}. The mean latency of LLC scans for the YCSB workload (Fig.~\ref{subfig:Scan_cycles}) is approximately 38 clock cycles. This latency is much lower than the number of LLC sets ($2K$ sets), resulting from the scope-buffer and SBV operations. To see the effectiveness of the SBV, Fig~\ref{subfig:sbv_skipratio} shows that the mean rate of skipped sets during a scan range is close to $94\%$ for the YCSB workload. For all TPC-H queries with all models, the mean SBV skipped sets ratio is above $99\%$ (not shown). In both Fig.~\ref{subfig:Scan_cycles} and Fig.~\ref{subfig:sbv_skipratio}, we see that our four models perform the same up to approximately $250$ scopes and then start to diverge. This divergence occurs when the back-pressure from the PIM module reaches the host memory controller and fills its buffers. At that point, PIM ops cannot enter the memory controller and the host cores must lower their PIM op issue rate. The lower issue rate allows PIM reads from other cores to interleave with the PIM ops, marking more cache sets in the SBV that are not skipped during an LLC scan. The stricter the model, the lower the issue rate becomes and the more interleaving it allows.

\begin{figure}[!t]
\centering

\begin{minipage}[c]{0.05\columnwidth}
\begin{subfigure}[c]{\textwidth}
\caption{}\label{subfig:inf_buf}
\end{subfigure}
\end{minipage}%
\begin{minipage}[c]{0.90\columnwidth}
\includegraphics[width=\textwidth,trim=0 0pt 0 0pt,clip]{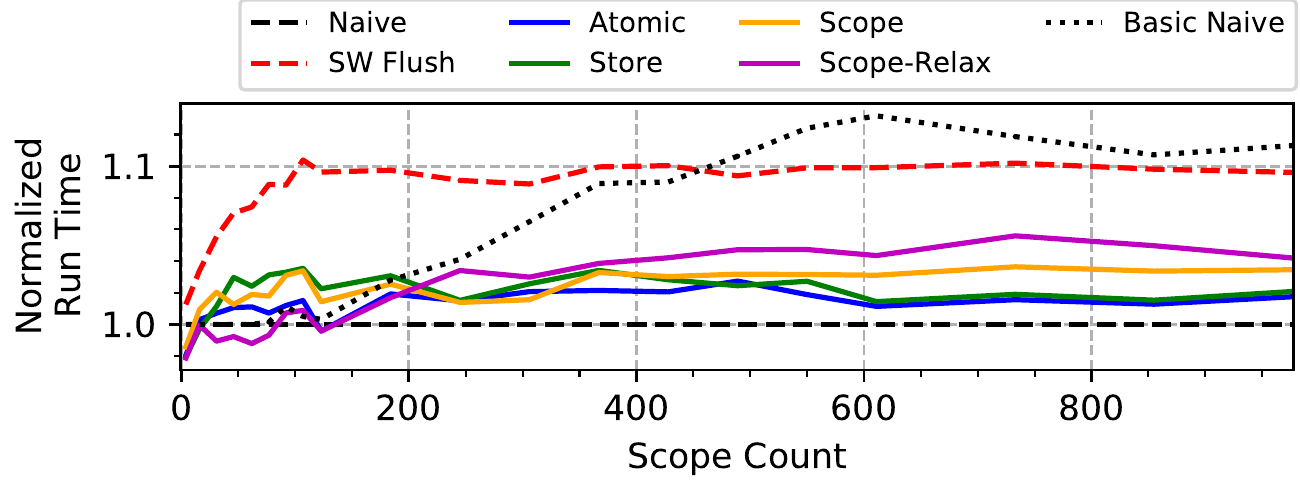}
\end{minipage}\\

\begin{minipage}[c]{0.05\columnwidth}
\begin{subfigure}[c]{\textwidth}
\caption{}\label{subfig:zero_logic}
\end{subfigure}
\end{minipage}%
\begin{minipage}[c]{0.90\columnwidth}
\includegraphics[width=\textwidth,trim=0 0pt 0 0pt,clip]{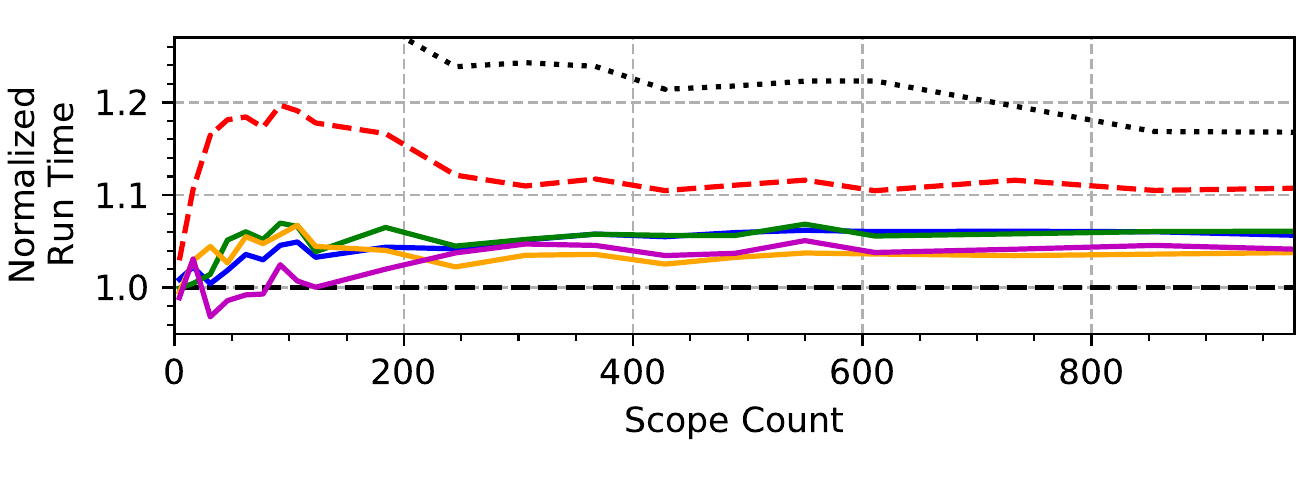}
\end{minipage}
\caption{Normalized run time for the YCSB workload and a PIM module with (a) an unbounded buffer and (b) zero PIM operation latency. Both (a) and (b) are normalized to the Naive baseline and also show the Naive baseline with the basic configuration (marked as \textit{Basic Naive}).}
\end{figure}

\textbf{Unbounded Buffer:} The limited buffer of the PIM module creates back-pressure and blurs the difference between the consistency models. To eliminate this effect, Fig.~\ref{subfig:inf_buf} shows the normalized run time for the YCSB workload when the PIM module buffer is unbounded. This measurement evaluates the system performance for the scenario where the PIM module buffer is sufficiently large for the used application. The unbounded buffer allows all PIM ops to reach the PIM module. Thus, operations to different scopes can be executed concurrently as soon as they reach the PIM module. Nevertheless, there is still no significant behavior difference between the consistency models because of the execution latency of the PIM module. The latter takes numerous cycles~\cite{PIMDB} and allows all consistency models to issue all PIM ops relatively quickly from the cores.

With an unbounded buffer, the Naive approach achieves the best run time. As PIM ops arrive to the memory at a faster rate, more parallelism in the PIM execution is uncovered. This benefit is small (less than $6\%$ relative to the four models) as the difference is in the PIM op management while the bulk of execution time is used for the PIM execution and read latency. Note that the scope and scope-relaxed models, though more relaxed and with the faster PIM op core issue rate, perform slightly worse ($2\%$) than the atomic and store models. The fast issue rate congests the host memory subsystem, hurting the execution of other memory operations as they pass through the same network and caches. The stricter models wait until a PIM op reaches the memory controller to release the next PIM op, with the side benefit being keeping the host memory subsystem clear. 

\textbf{Zero Logic:} To remove the effect of the PIM execution latency and account for a best-case bulk-bitwise PIM logic, we changed the basic configuration (Section~\ref{sec:methodology}) to have a bulk-bitwise PIM logic execution time of zero. The workload still issues the same PIM ops as before, but all PIM ops have a PIM execution time of zero. The experiment results for the YCSB workload, presented in Fig.~\ref{subfig:zero_logic}, show that the four models still have a maximum $6\%$ overhead compared to the Naive baseline. Additionally, the more relaxed consistency models, scope and scope-relaxed, achieve better performances. As the PIM execution time is zero, the system's management of PIM ops is the dominant factor for the PIM part of the workload; the faster issue rate of the relaxed models enables better performance.

\begin{figure}[!t]
\centering
\begin{minipage}[c]{0.05\columnwidth}
\begin{subfigure}[c]{\textwidth}
\caption{}\label{subfig:LLC_normalized}
\end{subfigure}
\end{minipage}%
\begin{minipage}[c]{0.90\columnwidth}
\includegraphics[width=\textwidth,trim=0 0pt 0 0pt,clip]{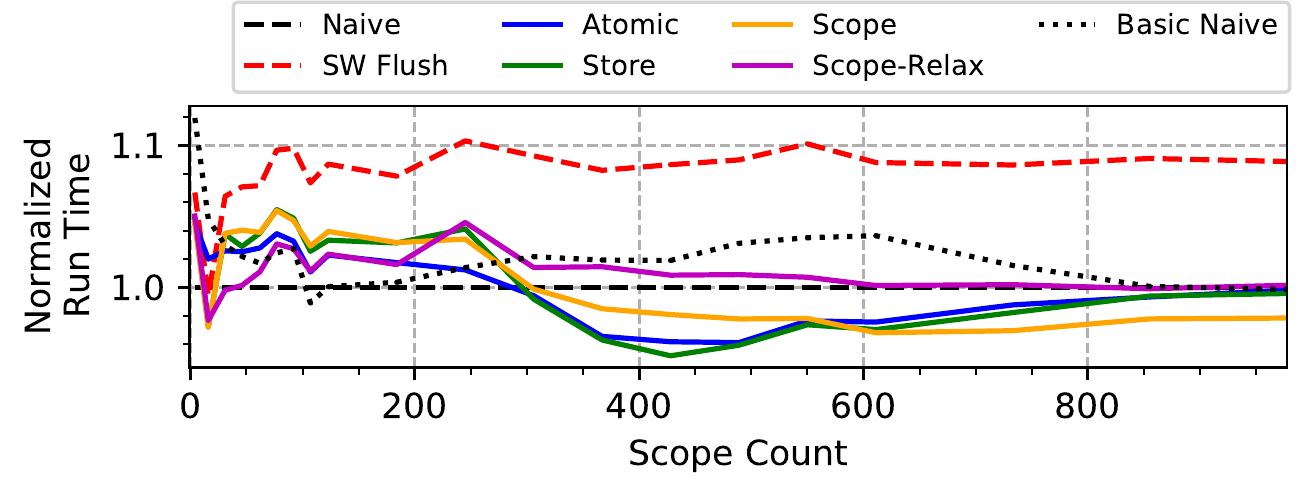}
\end{minipage}\\[-2pt]

\begin{minipage}[c]{0.05\columnwidth}
\begin{subfigure}[c]{\textwidth}
\caption{}\label{subfig:LLC_LLC_scan}
\end{subfigure}
\end{minipage}%
\begin{minipage}[c]{0.42\columnwidth}
\includegraphics[width=\textwidth,trim=0 7pt 0 0pt,clip]{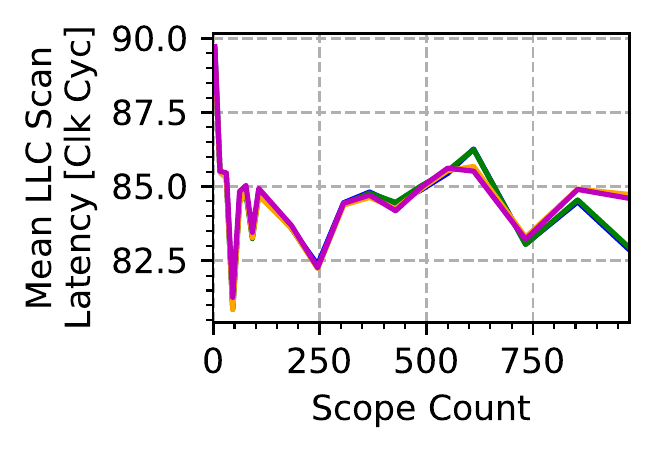}
\end{minipage}%
\begin{minipage}[c]{0.05\columnwidth}
\begin{subfigure}[c]{\textwidth}
\caption{}\label{subfig:LLC_LLC_SBV}
\end{subfigure}
\end{minipage}%
\begin{minipage}[c]{0.42\columnwidth}
\includegraphics[width=\textwidth,trim=0 7pt 0 0pt,clip]{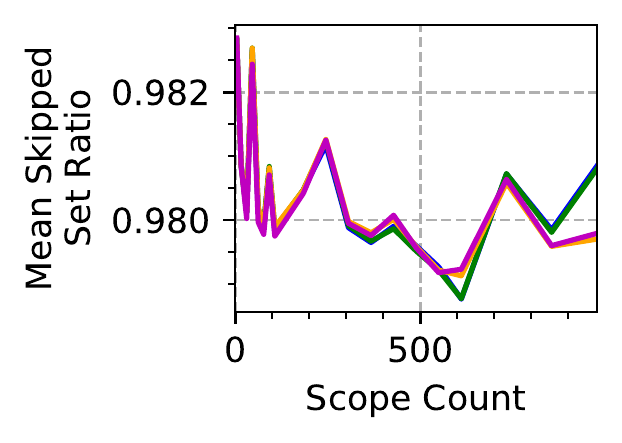}
\end{minipage}
\caption{Experiment result for configuration with an $8MB$ LLC for the YCSB workload. (a) Run time normalized to the Naive baseline, including the $2MB$ LLC Naive baseline (marked as \textit{Basic Naive}). (b) LLC scan mean cycle count. (c) SBV mean rate of LLC sets skipped during an LLC scan.
In (b) and (c), store overlaps atomic, and scope-relax overlaps scope. 
}
\label{fig:LLC_size}
\end{figure}

\textbf{LLC Size:} An interesting aspect is the effect of the LLC size on run time. A bigger LLC requires additional LLC scan latency and can increase the run time. Fig.~\ref{fig:LLC_size} shows the performance of our designs with an $8MB$ LLC for the YCSB workload. The behavior with the $8MB$ LLC (Fig.~\ref{subfig:LLC_normalized}) is similar to that with a $2MB$ LLC (Fig.~\ref{subfig:runtime_norm}), with the difference being a performance degradation relative to the Naive baseline. This degradation is due to the added LLC scan latency (Fig.~\ref{subfig:LLC_LLC_scan}), in spite of the increased SBV efficiency (Fig.~\ref{subfig:LLC_LLC_SBV}),  making the LLC busier and the host's memory subsystem more congested.

\begin{figure}[!t]
\centering
\includegraphics[trim=0 0 0 0pt, clip,width=\columnwidth]{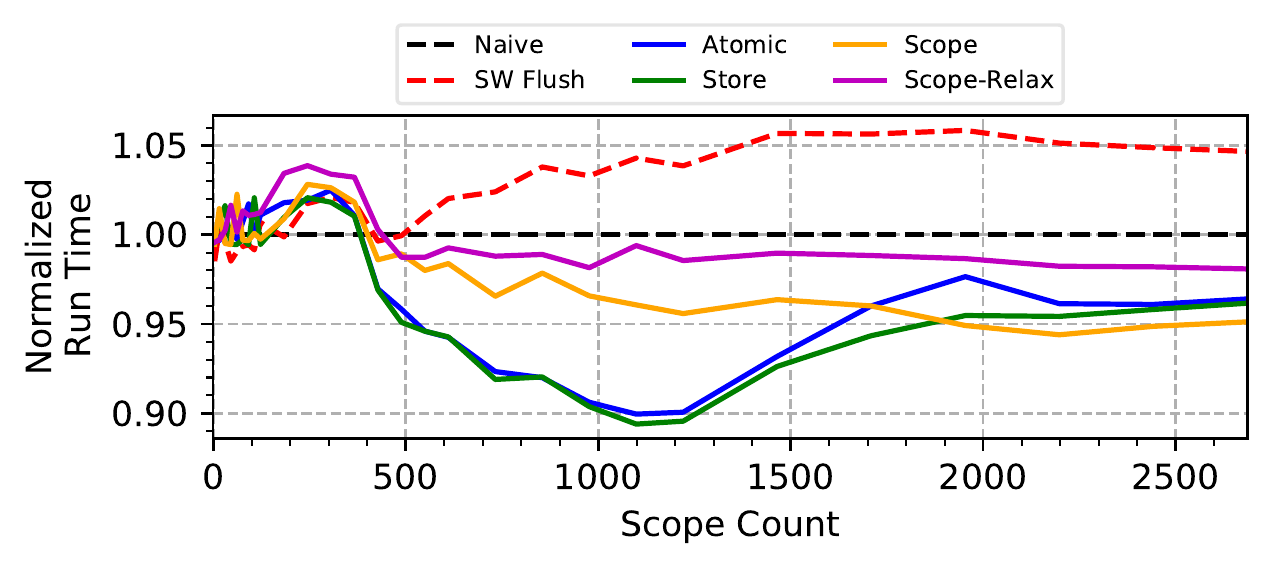}
\caption{Normalized run time for the YCSB workload using eight threads on a 16-core host processor. All models are normalized to the Naive baseline.}
\label{fig:threads}
\end{figure}

\textbf{Additional Threads:} Another interesting aspect is the effect of additional threads on the relative run time of the models. More threads might increase the load on the memory subsystem and increase the difference between the models.
Fig.~\ref{fig:threads} shows the performance of our designs for the YCSB workload where the scopes are divided between eight threads (instead of four). To allow all threads to run concurrently, we also increased the number of host cores to 16.
The behavior for eight threads shows the same trends as with four threads (shown earlier in Fig.~\ref{subfig:runtime_norm}), except that eight threads require scaling the scope count to achieve a similar scope count per thread. Additionally, as there are more threads and a greater load on the memory subsystem, the difference between the models and between the Naive baseline is even greater compared to with four threads, especially for the stricter atomic and store models. 

The above experiments show that even with the additional activities required to enforce order in our consistency models, PIM ops execution encounters greater bottlenecks: buffer congestion, PIM execution latency, and the result reads. The buffer congestion might be alleviated with a better design (\textit{e.g.}, larger PIM module buffer, better scheduling), but the PIM execution latency and the result read appear as inherent attributes of bulk-bitwise PIM~\cite{PIMDB,SIMDRAM,AMBIT,CONCEPT}, unrelated to the consistency model being used. 

\section{Related Work}
Previous work on bulk-bitwise PIM has mostly disregarded consistency, with some works addressing coherency. Perach \textit{et al.}~\cite{PIMDB} briefly mentioned that cores treat PIM operations as store operations (similar to our store model) but did not address the required support. In~\cite{PIMDB,SIMDRAM}, support for coherency is through explicit software flushes, which break ordering guarantees as described in Section~\ref{sec:Intro}. Seshadri \textit{et al.}~\cite{AMBIT} used the memory controller to flush cache-lines before PIM operations. To reduce the flush overhead, the memory controller flushes only cache-lines used by the PIM operation. Such a solution, as opposed to our coherency solution, assumes that the memory controller knows what cache-lines the PIM operations use, making the host and PIM module tightly integrated, which can be impractical in a general scenario.

Prior work on PIM consistency models and coherency for near-memory architecture~\cite{HMC} does exist~\cite{Lee2015,Liu2018,LazyPIM,Ahn2015,OrderLight}. Near-memory locates processing units on the same die as the memory arrays, but the processing units and memory arrays are distinct and separated modules. As near-memory computing and bulk-bitwise PIM differ substantially in implementation and supported operations, they are used in different ways and suited to different computations. Hence, these solutions for near-memory are mostly not appropriate for bulk-bitwise PIM.
In a work on near-memory computation that is relevant for bulk-bitwise PIM, Nag \textit{et al.}~\cite{OrderLight}, showed that standard memory fence operations are insufficient to enforce order among PIM operations. They suggested a new fence mechanism for that goal and demonstrated it on a near-memory architecture with a GPU host. Nag \textit{et al.}, however, did not discuss the order of PIM operations concerning other memory operations and thus did not provide a consistency model. We use their fence mechanism as our fence operation for PIM ops in our proposed models (Section~\ref{sec:consistency_support}). 
\section{Conclusions}
In this paper, we showed the importance of addressing the consistency and coherency of bulk-bitwise PIM systems. We proposed four bulk-bitwise PIM consistency models, from strict to relaxed, and discussed the implementation to support these models, including a low hardware overhead solution for coherency (the scope buffer and the SBV). These consistency models were evaluated on representative database workloads. Our evaluation showed that strict and relaxed models for bulk-bitwise PIM can have similar run times, which are also similar to the run time of a system with no order guarantees. This is because the bulk-bitwise PIM bottlenecks overshadow the overheads associated with the consistency models. Nevertheless, a consistency model and implementation that take into account these bottlenecks -- such as our scope model that inherently interleaves PIM operations and lightly loads the host memory subsystem -- can reach better performance in some cases and might be a better choice for bulk-bitwise PIM.

%%%%%%% -- PAPER CONTENT ENDS -- %%%%%%%%

%%%%%%%%% -- BIB STYLE AND FILE -- %%%%%%%%
\bibliographystyle{IEEEtranS}
\bibliography{refs}
%%%%%%%%%%%%%%%%%%%%%%%%%%%%%%%%%%%%

\end{document}